\documentclass[aps,prl,twocolumn,showpacs]{revtex4}

\usepackage{amsmath}
\usepackage{graphicx}

\begin{document}

\title{Cavity assisted quasiparticle damping in a Bose-Einstein condensate}
\author{S.A.~Gardiner,$^{1,2,3}$ K.M.~Gheri,$^{1}$ and P.~Zoller$^{1}$}
\affiliation{$^{1}$Institut f{\"u}r Theoretische Physik,
Universit{\"a}t Innsbruck, A--6020 Innsbruck, Austria\\
$^{2}$Institut f\"{u}r Physik, Universit\"{a}t Potsdam,
D--14469 Potsdam, Germany\\
$^{3}$Institut f{\"u}r Theoretische Physik,
Universit{\"a}t Hannover, D--30167 Hannover, Germany
}

\begin{abstract}
We consider an atomic Bose-Einstein condensate held within an optical cavity and
interacting with laser fields. 
We show how the interaction of the cavity
mode with the condensate can cause energy due to excitations 
to be coupled to
a lossy cavity mode, which then decays, 
thus damping the condensate, 
how to
choose parameters for damping specific excitations, 
and how to
target a range of different excitations to 
potentially 
produce extremely cold  
condensates.
\end{abstract}

\maketitle

The breakthrough success,
largely due to improved cooling techniques \cite{cooling},
of producing atomic Bose-Einstein condensates (BEC) \cite{experiments},
has created an exciting new area of investigative study \cite{excitingBEC,
stringari}.
Within the field of cavity quantum electrodynamics, one is able to
precisely control the interaction of the electromagnetic field with an
atom \cite{excitingcavity}; the trapping and cooling of a single atom in an optical cavity has
been successfully experimentally demonstrated \cite{singleatomcool}. Here we
combine \cite{combine} the use a cavity to cool 
atoms \cite{coolfew}
with the ever present goal of producing colder condensates 
\cite{horak}, particularly relevant to controlling 
quasiparticle excitations \cite{bogoliubov,numberbog,numberbog2}
produced in the course of other
experimental investigations \cite{phase}.
\begin{figure}
\begin{center}
\includegraphics[width=45mm]{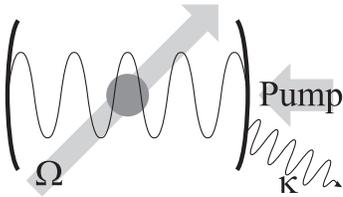}
\caption{Proposed configuration: a BEC held in a cavity, driven by a
laser $\Omega$ and a pump laser, and
losing cavity photons with a decay rate $\kappa$.}
\label{cavitypic}
\end{center}
\end{figure}

We consider a Fabry-Perot optical cavity with decay rate
$\kappa$, and
a pump laser, with a BEC held in a harmonic potential of frequency $\omega$
inside the cavity, and driven by a separate
laser $\Omega$, as shown in Fig.~\ref{cavitypic}. 
We generally
consider only one spatial dimension, considering the radial motion 
to be frozen out by a tight harmonic potential of frequency $\omega_{r}$
(cigar-shaped configuration \cite{oneD}). 
The laser-cavity superposition field is
$\hat{E}(\hat{x},t)=\Omega(\hat{x})e^{-i\omega_{L}t}+g\hat{a}\cos(k\hat{x})$
where $\hat{a}$ is the cavity mode annihilation operator;
interacting with this are the BEC atoms,
considered to have two levels $|g\rangle$ and $|e\rangle$, with
transition frequency $\omega_{0}$.
We initially consider a single particle, and ignore the
degrees of freedom due to the free cavity field and the atomic motion.
In a rotating frame 
($\hat{U}=\exp
[-i(\sigma_{ee}+\hat{a}^{\dagger}\hat{a})\omega_{L}t]$),
the Hamiltonian describing the internal atomic level dynamics is
$\hat{H}_{\rm at} =\hbar[\Delta_{a}\sigma_{ee}+
\tilde{E}(\hat{x})\sigma_{eg}+
\tilde{E}^{\dagger}(\hat{x})\sigma_{ge}]$,
where $\sigma_{eg}=|e\rangle\langle g|$, $\Delta_{a}=\omega_{0}-\omega_{\mbox{\scriptsize ex}}$,
and $\tilde{E}(\hat{x})=\Omega(\hat{x})+g\hat{a}\cos(k\hat{x})$.
Assuming $\Delta_{a}\gg \tilde{E}$, $\Delta_{a}\gg \gamma$ where $\gamma$ is 
the atomic spontaneous emission rate, 
and $|\Delta|$, 
where $\Delta=\omega_{c} -\omega_{\mbox{\scriptsize ex}}$, by far
exceeding the frequency scales governing
atomic motion and the cavity dynamics, we may adiabatically eliminate
$|e\rangle$ 
to derive 
a closed time-evolution 
equation
for the $|g\rangle$ wavefunction component
\cite{quantumnoise}, using this to reconstruct an effective Hamiltonian:
\begin{equation}
\hat{H}_{\rm eff}=-\frac{\hbar\Delta_{a}}{\gamma^{2}+\Delta_{a}^{2}}
\tilde{E}^{\dagger}(\hat{x})\tilde{E}(\hat{x})
+\frac{i\hbar\gamma}{\gamma^{2}+\Delta_{a}^{2}}
\tilde{E}^{\dagger}(\hat{x})\tilde{E}(\hat{x}).
\label{Heffend}
\end{equation}
Assuming $g\ll \Omega$ and $\langle \hat{a}^{\dagger}\hat{a}\rangle \le 1$,
we can neglect the terms $\propto g^{2}$, and as 
$\gamma\ll\Delta_{a}$, we ultimately consider only the
Hermitian terms. 
The light field thus provides:
a term $\propto |\Omega(\hat{x})|^{2}$, which, assuming $\Omega(\hat{x})\propto
e^{ik_{L}\hat{x}}$, contributes only a
shift; and a contribution linear in $\hat{a}$ and $\hat{a}^{\dagger}$, which 
acts like a driving field for the cavity mode. It thereby implements a mechanism to transfer energy to
the cavity which in turn will shed this energy to
the environment by means of cavity damping.
The effective single particle Hamiltonian (including the free dynamics of
the particle motion and the cavity) is now
\begin{equation}
\hat{H}_{\mbox{\scriptsize eff}}= 
\frac{\hat{p}^{2}}{2m}+\frac{m\omega^{2}\hat{x}^{2}}{2m}+
\hbar
(\Delta
-i\kappa)\hat{a}^{\dagger}\hat{a}+
\hbar
[f(\hat{x})\hat{a}+\mbox{H.c.}],
\end{equation}
where
$f(\hat{x})=-\Omega(\hat{x})g\cos(k\hat{x})\Delta_a/(\Delta_a^2+\gamma^2 )$,
and $m$ is the particle mass. 
We now consider  a BEC-cavity system.
The one-dimensional atomic interaction potential
$=u\delta(x-y)$, where $u=2\hbar\omega_{r}a_{s}$ and 
$a_{s}$ is the $s$-wave scattering length \cite{stringari}.
Thus, in second quantized form:
\begin{multline}
\hat{H}_{\mbox{\scriptsize eff}}=\int dx
\hat{\Psi}^{\dagger}(x)\biggl\{
\hbar \left[f(x)\hat{a} +f^{*}(x)\hat{a}^{\dagger}\right]
\\
-\frac{\hbar^{2}}{2m}\frac{\partial^{2}}{\partial x^{2}}
+\frac{m\omega^{2}x^{2}}{2}
+ \frac{u}{2}\hat{\Psi}^{\dagger}(x)
\hat{\Psi}(x)
\biggr\}\hat{\Psi}(x)
\\
+\hbar\left[(\Delta-i\kappa)\hat{a}^{\dagger}\hat{a}
-\alpha_{p}\hat{a}^{\dagger}-\alpha_{p}^{*}\hat{a}\right],
\end{multline}
where $\hat{\Psi}(x)$ and $\hat{\Psi}^{\dagger}(x)$ are
atomic field operators, and the $\alpha_{p}$ terms are provided by the
pump laser.
We treat
$\hat{\Psi}(x)$ and $\hat{a}$ semiclassically, replacing them with the scalar
quantities
$\varphi(x)$ and $\alpha$, respectively. 
It is convenient
to rescale to dimensionless harmonic units $(\hbar = m = \omega = 1)$;
this scaling is assumed from now on.
The resulting equations of motion are: a
Gross-Pitaevskii-like (GP) equation
\begin{equation}
i\dot{\varphi}(x)=\{H_{\mbox{\scriptsize GP}}
+[\tilde{f}(x)\tilde{\alpha}
+\tilde{f}^{*}(x)\tilde{\alpha}^{*}]\}\varphi(x),
\label{keyeqnone}
\end{equation}
where 
$H_{\mbox{\scriptsize GP}}=-\partial^{2}/2\partial x^{2}
+x^{2}/2 + \upsilon|\varphi(x)|^{2}$ 
is the unperturbed GP
``Hamiltonian,'' with 
$\upsilon = uN\sqrt{m/\hbar\omega}/\hbar$,
$\tilde{f}(x)=\sqrt{N}f(x)/\omega$, and $\tilde{\alpha}=\alpha/\sqrt{N}$, 
where
$N$ is the particle number;
and the semiclassical cavity mode equation
\begin{equation}
i\dot{\tilde{\alpha}}=(\Delta-i\kappa)\tilde{\alpha}+\int dx |\varphi(x)|^{2}
\tilde{f}^{*}(x)
-\tilde{\alpha}_{p}.
\label{keyeqntwo}
\end{equation}
We set $\tilde{\alpha}_{p}$ to be
$= \int dx |\varphi_{0}(x)|^{2}\tilde{f}^{*}(x)$,
so that when $\varphi(x)$ approaches $\varphi_{0}(x)$
(the ground state of $H_{\mbox{\scriptsize GP}}$),
$\tilde{\alpha}$ will simply
decay, without feeding into Eq.~(\ref{keyeqnone}). The 
cooling mechanism
thus switches off upon reaching the desired steady state 
$\varphi(x)=\varphi_{0}(x)$ and $\tilde{\alpha} = 0$.

\begin{figure}
\begin{center}
\includegraphics[width=80mm]{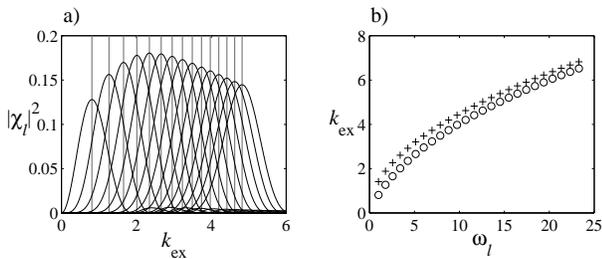}
\caption{a) Plots of $|\chi_{l}|^{2}$ for $l=$1--15 as functions of
$k_{\mbox{\scriptsize ex}}$, where $\chi_{l}$ is as defined
in Eq.~(\ref{chidef}), and $\tilde{f}(x)=e^{ik_{\mbox{\tiny ex}}x}$. The
vertical lines mark the numerically determined optimal values for
$k_{\mbox{\scriptsize ex}}$.
b) Plots of: $k_{\mbox{\scriptsize ex}}=\sqrt{2\omega_{l}}$, i.e.\ approximately
optimal value of $k_{\mbox{\scriptsize ex}}$
(plusses), and numerically determined optimal vales of 
$k_{\mbox{\scriptsize ex}}$ (circles),
against $\omega_{l}$ for $l=$1--25.}
\label{peaks}
\end{center}
\end{figure}

We now consider linearized perturbations $\delta \varphi(x)$, $\delta\alpha$,
around the steady state.
Linearizing
Eq.~(\ref{keyeqnone}) produces Bogoliubov-like equations \cite{bogoliubov}.
Note in Eq.~(\ref{keyeqntwo}) for $\varphi_{0}(x)$ no distinction 
is made between different global phases; in calculations 
the final phase of $\varphi_{0}(x)$ is 
determined by the chosen initial conditions, and is not relevant. 
We 
consider only  perturbations orthogonal to $\varphi_{0}(x)$, and can
implicitly assume a $U(1)$ gauge transformation \cite{gauge} such that  
$\varphi_{0}(x)$  is always real.
Having defined $\delta\varphi(x)$ as orthogonal to $\varphi_{0}(x)$, i.e.\
$\delta\varphi(x) = \int dy Q(x,y)\delta\varphi(y)$ where $Q(x,y) =
\delta(x-y)-\varphi_{0}(x)\varphi_{0}(y)$, we can state that
\begin{multline}
i\left(
\begin{array}{c}
\delta\dot{\varphi}(x)\\
\delta\dot{\varphi}^{*}(x)
\end{array}
\right)
=
\int dy {\mathcal L}(x,y)
\left(
\begin{array}{c}
\delta\varphi(x)\\
\delta\varphi^{*}(x)
\end{array}
\right)
\\
+\int dy {\mathcal Q}(x,y)
[\tilde{f}(y)\delta\alpha+\tilde{f}^{*}(y)\delta\alpha^{*}]
\left(
\begin{array}{c}
\varphi_{0}(y)\\
-\varphi_{0}^{*}(y)
\end{array}
\right),
\label{theequation}
\end{multline}
where ${\mathcal L}(x,y) =
\int\int dz dw{\mathcal Q}(x,z)H_{\mbox{\scriptsize Bog}}(z,w)
{\mathcal Q}(w,y)$ \cite{numberbog2},
in terms of the usual Bogoliubov
Hamiltonian \cite{bogoliubov}:
\begin{equation}
H_{\mbox{\scriptsize Bog}}=
\left(
\begin{array}{cc}
H_{\mbox{\scriptsize GP}} + \upsilon|\varphi_{0}(x)|^{2} -\mu& 
\upsilon\varphi_{0}(x)^{2}\\
-\upsilon\varphi_{0}^{*}(x)^{2} & -H_{\mbox{\scriptsize GP}} -  
\upsilon|\varphi_{0}(x)|^{2} +\mu
\end{array}
\right)
\end{equation}
where $\mu$ is the ground state chemical potential; 
and 
\begin{equation}
{\mathcal Q}(x,y) =
\left(
\begin{array}{cc}
Q(x,y)& 0\\
0 & Q(x,y)
\end{array}
\right).
\end{equation}
It is convenient to expand $(\delta\varphi (x),\delta\varphi^{*} (x))$ as:
\begin{equation}
\left(
\begin{array}{c}
\delta\varphi (x) \\
\delta\varphi^{*} (x) 
\end{array}
\right)=
\sum_{k}
\zeta_{k}
\left(
\begin{array}{c}
u_{k} (x) \\
v_{k} (x) 
\end{array}
\right)+
\zeta_{k}^{*}
\left(
\begin{array}{c}
v_{k} (x) \\
u_{k} (x) 
\end{array}
\right),
\label{stateexpand}
\end{equation}
where $\{(u_{k}(x),v_{k}(x)),(v_{k}(x),u_{k}(x))\}$ are eigenstates of
${\mathcal L}(x,y)$, with eigenfrequencies $\pm
\omega_{k}$ (these may be determined numerically by diagonalizing ${\mathcal
L}(x,y)$ \cite{me}). As $\varphi_{0}$ is assumed real, these are
also real. We similarly expand: 
\begin{multline}
\int dy {\mathcal Q}(x,y)
\tilde{f}(y)
\left(
\begin{array}{c}
\varphi_{0}(y)\\
-\varphi_{0}(y)
\end{array}
\right)
=
\\
\sum_{k}
\chi_{k}
\left[
\left(
\begin{array}{c}
u_{k}(x)\\
v_{k}(x)
\end{array}
\right)-
\left(
\begin{array}{c}
v_{k}(x)\\
u_{k}(x)
\end{array}
\right)
\right],
\label{bogmatch}
\end{multline}
where the coefficients can obviously be defined by
\begin{equation}
\chi_{k} = \int dx [u_{k}(x) + v_{k}(x)]\tilde{f}(x)\varphi_{0}(x).
\label{chidef}
\end{equation}
Note that treating the evolution of perturbations 
orthogonal to $\varphi_{0}(x)$, is equivalent to 
a number-conserving formalism \cite{numberbog,numberbog2}.
The $u_{k}(x)$ and $v_{k}(x)$ are orthogonal to $\varphi_{0}(x)$, and the
$\{(u_{k}(x),v_{k}(x)),(v_{k}(x),u_{k}(x))\}$ are used as a convenient
time-independent basis.
All time dependence in Eq.~(\ref{stateexpand}) is
thus in the $\zeta_{k}$, $\zeta_{k}^{*}$ coefficients, in contrast
to Refs.~\cite{numberbog,numberbog2,gauge}.
We now transform Eq.~(\ref{theequation}) to an interaction picture. 
We thus set $(\delta\tilde{\varphi}(x),\delta\tilde{\varphi}^{*}(x))
=e^{i{\mathcal L}t}
(\delta \varphi (x),\delta \varphi^{*}(x))$ and
$\delta \tilde{\alpha} = e^{i\Delta t}\delta \alpha$.
Substituting
Eqs.~(\ref{stateexpand}) and (\ref{bogmatch}) into an appropriately transformed 
Eq.~(\ref{theequation}),
and then making the integration
$\int dx u_{l}(x)\delta\dot{\tilde{\varphi}}(x) -
v_{l}(x)\delta\dot{\tilde{\varphi}}^{*}(x)$, we end up with
\begin{equation}
i\dot{\tilde{\zeta}}_{l} = 
\delta\tilde{\alpha}\chi_{l}
e^{i(\omega_{l}-\Delta)t} +
\delta\tilde{\alpha}^{*}\chi_{l}^{*}
e^{i(\omega_{l}+\Delta)t},
\label{fundamental1}
\end{equation}
where $\tilde{\zeta}_{l} = e^{i\omega_{l}t}\zeta_{l}$. 
From Eqs.~(\ref{keyeqntwo}) and (\ref{bogmatch}), 
the linearized equation of motion for $\delta\tilde{\alpha}$, after
adiabatic elimination 
(assuming $\kappa \gg
|\beta_{k}\chi_{k}|$), produces 
\begin{equation}
\delta\tilde{\alpha} = -\frac{i}{\kappa}
\sum_{k}[
\tilde{\zeta}_{k}e^{-i(\omega_{k}-\Delta)t}+
\tilde{\zeta}_{k}^{*}e^{i(\omega_{k}+\Delta)t}
]\chi_{k}^{*}.
\label{cavmatch}
\end{equation}
We assume that due to our choice of $\tilde{f}(x)$, 
$|\chi_{l}|$ dominates $|\chi_{k\neq l}|$,
neglecting terms where $k \neq l$, and choose 
$\Delta = \omega_{l}$ to match it.
Applying the rotating wave approximation (RWA) to
Eqs.~(\ref{cavmatch}) and (\ref{fundamental1}), we
neglect terms involving $\tilde{\zeta}_{l}^{*}$ and $\delta\tilde{\alpha}^{*}$
($2\omega_{l}$ should therefore describe the fastest timescale). Combining the
resulting expressions,
we end up 
with a simple damping equation:
\begin{equation}
\dot{\tilde{\zeta}}_{l} = -\frac{
|\chi_{l}|^{2}}{\kappa}\tilde{\zeta}_{l}.
\label{bogdamp}
\end{equation}
As $\Omega(x)\propto e^{ik_{L}x}$, we set 
$\tilde{f}(x)=g_{0}[e^{ik_{\mbox{\tiny ex}}x}
+e^{i(k_{\mbox{\tiny ex}}\pm 2k)x}]$, where $k_{\mbox{\scriptsize ex}}=k\pm
k_{L}$, and $k$ is the wavenumber of the cavity mode. 
Taking only the $e^{ik_{\mbox{\tiny ex}}x}$ term, we 
see in Fig.~\ref{peaks}b) that each 
$|\chi_{l}|^{2}$
peaks for some $k_{\mbox{\scriptsize ex}}$ at a point 
where $|\chi_{l}|^{2}>|\chi_{k\neq l}|^{2}$. Assuming
$\omega_{l}$ is chosen to match,
for sufficiently large $k$ the $e^{i(k_{\mbox{\tiny ex}}\pm 2k)x}$ term can
be ignored, as those modes to which it couples most
strongly (i.e.\ large $\chi_{l}$) can, by the RWA, be neglected. 
From now on we consider 
$\tilde{f}(x)=g_{0}e^{ik_{\mbox{\tiny ex}}x}$. 
As  $e^{ik_{\mbox{\tiny
ex}}x}\varphi_{0}(x)$ is equivalent to a momentum-kicked ground state 
with mean additional energy $=k_{\mbox{\scriptsize ex}}^{2}/2$, 
the inner product of
$\tilde{f}(x)(\varphi_{0}(x),-\varphi_{0}(x))$
with eigenstates of ${\mathcal L}(x,y)$ should be large with those eigenstates
of similar energy; 
one would thus simplistically 
expect $|\chi_{l}|^{2}$ to 
peak when $k_{\mbox{\scriptsize ex}}^{2}/2=\omega_{l}$.
In Fig.~\ref{peaks}b) we see that this prediction is a little 
crude for small $\omega_{L}$, although it converges to the true values
for higher frequencies, as expected. Also correctly predicted is that the optimal
$k_{\mbox{\scriptsize ex}}$ values tend to converge for large $l$.
From Fig.~\ref{peaks}a) we see that the
$|\chi_{l\pm 1}|^{2}$ (at least) can be significant compared to 
$|\chi_{l}|^{2}$ at the optimal value of $k_{\mbox{\scriptsize ex}}$, but the 
fact that these terms are both smaller and rotating means that the $l$th term will
still dominate. For large $l$, as the
peak values of $k_{\mbox{\scriptsize ex}}$ converge and the peaks
become less well resolved, Eq.~(\ref{bogdamp}) will 
become less reliable.

\begin{figure}
\begin{center}
\includegraphics[width=80mm]{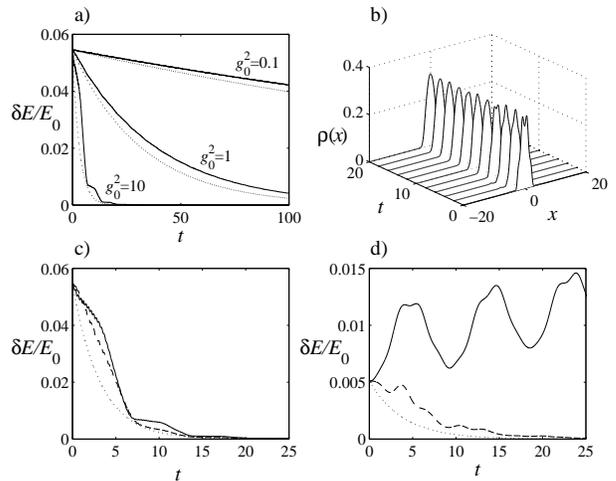}
\caption{
a) Plots of $\delta E/E_{0}$ for different $g_{0}^{2}$ against time, 
where $l=12$, $\kappa = 10$.  
Solid lines, numerical calculations; 
dotted lines, prediction of Eq.~(\ref{energydamp}).  
b) Time evolution of $\rho(x)=|\varphi(x)|^{2}$ 
($l=12$, $\kappa$, $g_{0}^{2}=10$). 
c) Plots of $\delta E/E_{0}$ against time, where $l=12$.
Solid line, $\kappa$, $g_{0}^{2}=10$;
dashed line, $\kappa$, $g_{0}^{2}=1$; 
dotted line, prediction of Eq.~(\ref{energydamp}) for both cases. 
d) Plots of $\delta E/E_{0}$ against time, where $l=1$.
Solid line, $\kappa$, $g_{0}^{2}=10$;
dashed line, $\kappa$, $g_{0}^{2}=1$; 
dotted line, prediction of Eq.~(\ref{energydamp}).
}
\label{onemode}
\end{center}
\end{figure}

Numerically we monitor the damping via the 
energy 
$E = \int dx 
\varphi^{*}(x)
[-\partial^{2}/\partial x^{2}
+ x^{2} +\upsilon|\varphi(x)|^{2}
]\varphi(x)/2$. 
We define $E_{0}$ equivalently, with
$\varphi(x)=\varphi_{0}(x)$. Within the linearized approximation 
$E = E_{0} + \sum_{k}\omega_{k}|\zeta_{k}|^{2}$,
and thus, if initially only $\zeta_{l}\neq 0$, from Eq.~(\ref{bogdamp}) we 
have 
\begin{equation}
\delta\dot{E} = -\frac{2|\chi_{l}|^{2}}{\kappa}\delta E,
\label{energydamp}
\end{equation}
where $\delta E = E - E_{0}$.
The energy damping timescale is thus $=2g_{0}^{2}A/\kappa$, where $A =
|\chi_{l}|^{2}$ for $g_{0}=1$. In Fig.~\ref{onemode} we compare
Eq.~(\ref{energydamp}) with numerical
integrations of Eqs.~(\ref{keyeqnone}) and (\ref{keyeqntwo}). The initial
conditions are $\varphi(x) \propto \varphi_{0}(x) + 
0.1[u_{l}(x)+v_{l}(x)]$, $\alpha=0$; $\upsilon$ is always $=10$.
In Fig.~\ref{onemode}a) we see qualitative agreement of the analytical
estimate given by Eq.~(\ref{energydamp}) with the numerical results for 
different $g_{0}$, which, in view of the number of approximations made, is
remarkably good. In Fig.~\ref{onemode}b) we see the convergence of the
position density $\rho(x)=|\varphi(x)|^{2}$ to $|\varphi_{0}(x)|^{2}$. In
Fig.~\ref{onemode}c) we see good qualitative agreement in the damping rates for
$g_{0}^{2}$ and $\kappa=1$, $10$ with the analytical prediction
(which is the same for each). In Fig.~\ref{onemode}d), where $l=1$ rather
than 12, but the
parameters are otherwise the same, we see heating for
the $\kappa = 10$ case; $\omega_{1}=1$ is dominated by
$\kappa$, and the RWA is therefore not valid. 
Most significant about equations 
(\ref{bogdamp}) and (\ref{energydamp}) is that 
the procedure for deriving them provides a recipe for determining parameters
optimal for the damping of specific excitations. 

\begin{figure}
\begin{center}
\includegraphics[width=80mm]{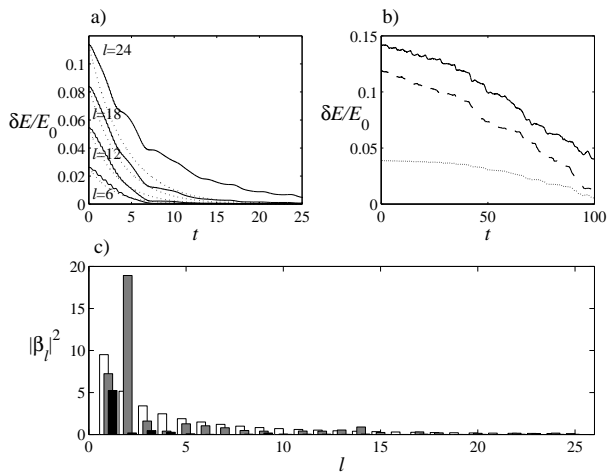}
\caption{
a) Damping of states perturbed by a single excitation, for 
$\kappa$, $g_{0}^{2}=1$.
Solid lines, numerical calculation; dotted lines, analytical estimates.
b) Plots of $\delta E/E_{0}$ against time:
solid line, $\tau=10$; dashed line, same initial condition, but with
a cutoff for $l>25$; dotted line, $\tau=5$.
c) White, $\langle \hat{b}_{l}^{\dagger}\hat{b}_{l}\rangle$
for $\tau=10$. Data
corresponding to the solid line of b):
grey,  $|\beta_{l}|^{2}$ for $t=0$; 
black,  $|\beta_{l}|^{2}$ for $t=100$.
}
\label{thermaldamp}
\end{center}
\end{figure}

In the case of a finite
temperature BEC, where there is a range of populated quasiparticle (QP)
excitations, these can be individually targeted, resulting in cooling of the
BEC.
There is a well known correspondence between linearized perturbations of 
$\varphi(x)$
and QP excitations \cite{numberbog,numberbog2}.
Assuming
$\hat{a}_{0}^{\dagger}\hat{a}_{0}\approx \hat{N}$, 
where
$\hat{a}_{0}$ is the annihilation operator for a particle in state
$\varphi_{0}(x)$ and $\hat{N}$ is
the total particle number
operator, then
\begin{equation}
\hat{\Psi}(x)=\hat{a}_{0}\left[\varphi_{0}(x)+\frac{1}{\sqrt{\hat{N}}}
\sum_{k}\hat{b}_{k}u_{k}(x)+\hat{b}_{k}^{\dagger}v_{k}(x)\right],
\end{equation}
where the $\hat{b}_{k}^{\dagger}$, $\hat{b}_{k}$ are QP creation and 
annihilation operators, respectively. If the state
of the system is assumed thermal, then
$\langle\hat{b}_{k}^{\dagger}\hat{b}_{k}\rangle 
=[\exp(\omega_{k}/\tau)-1]^{-1}$ and $\langle \hat{b}_{k}\rangle = 
\langle \hat{b}_{k}^{\dagger}\rangle = 0$, where $\tau = k_{B}T/\hbar\omega$.
Semiclassically one can regard these expectation values as describing the
statistics of  Gaussian random variables $\beta_{k}$, with
mean 0 and variance 
$\langle\hat{b}_{k}^{\dagger}\hat{b}_{k}\rangle$ \cite{wigners}.
One can make use of the linearized perturbation-QP correspondence
by taking a random initial condition:
\begin{equation}
\varphi(x) = \varphi_{0}(x) +
\frac{1}{\sqrt{N}}\sum_{k}\beta_{k}u_{k}(x)+\beta_{k}^{*}v_{k}(x),
\label{initial}
\end{equation}
with $\alpha=0$, and simulating the cooling process
numerically with Eqs.~(\ref{keyeqnone}), (\ref{keyeqntwo}). 
As we must target a range of excitations,
we employ a procedure so that
whenever $t$ is a multiple of 4, $k_{\mbox{\scriptsize ex}}$ and $\Delta$ are 
changed (by adjusting the laser $\Omega$), initially targeting 
$l=25$ (as described above), and then through each of $l=$24--1. We choose 
$g_{0}^{2}$, $\kappa=1$ 
(to ensure cooling takes place for small $\omega_{l}$), $N=1000$, and
$\upsilon = 10$. 
In Fig.~\ref{thermaldamp}a) we show damping of states of the form 
$\varphi(x)\propto \varphi_{0}(x)+0.1[u_{l}(x)+v_{l}(x)]$, for a range of
relevant $l$. 
For $l=24$ there are deviations from Eq.~(\ref{energydamp}), as expected;
significant damping nevertheless takes place for each $l$. In
Fig.~\ref{thermaldamp}b) we observe significant damping for initial states
where $\tau=5$, 10. Note that the energy
separation between the $\tau=10$ state, and the ``cutoff'' 
state [$k=1$--25 in  Eq.~(\ref{initial}), rather than 1--100], 
remains relatively static, implying that damping is
taking place in a targeted manner on the low-lying excitations. 
Note also from Fig.~\ref{thermaldamp}c) that although the relative population for
$k>25$ is miniscule, the relative energy contribution is significant. 
Although
the cutoff state was propagated for the sake of comparison, it could
conceivably apply to a physical situation, such as occurs in evaporative
cooling \cite{cooling}. In Fig.~\ref{thermaldamp}c) we see clear supression of 
the $|\beta_{l}|^{2}$, where $\beta_{l}$ is determined by 
$\sqrt{N}\int dx \varphi^{*}(x)u_{l}(x) -\varphi(x)v_{l}(x)$. By our
semiclassical analogy, it follows that QP populations will similarly
decrease,  
potentially providing a method to obtain extremely cold
condensates. Note that some spontaneous
emission is inevitable, and the resulting heating must be slower than
the overall cooling rate. Details on the
effect of spontaneous emission on BEC temperature are in
Ref.~\cite{fedichev}.

We have presented a cavity-laser-BEC configuration whereby, through astute
choice of parameters, specific excitations of the BEC can be rapidly damped.
We have shown the derivation of a simple equation which explains all major
features of the damping process, and provides a recipe for determining optimal
parameters. We have demonstrated how this damping procedure can in priniciple be
used to produce extremely cold condensates.

We thank 
R. Dum, 
T. Felbinger, 
C.W. Gardiner,
D. Jaksch, 
P. Horak, 
and H. Ritsch for
discussions, the Austrian Science
Foundation, and the European Union TMR network ERBFMRX-CT96-0002.
S.A.G. also thanks R. Gardiner,
the European Science Foundation, and the Alexander von Humboldt Foundation.

\end{document}